# Computational Strategy for Graphene: Insight from Odd Electrons Correlation

E. F. Sheka

Peoples' Friendship University of Russia, Miklukho-Maklay, 6, Moscow 117198, Russia

sheka@icp.ac.ru

**Abstract.** The correlation of odd electrons in graphene turns out to be significant so that the species should be attributed to correlated ones. This finding profoundly influences the computational strategy addressing it to multireference computational schemes. Owing to serious problems related to the schemes realization, a compromise can be suggested by using single-determinant approaches based on either Hartree-Fock or Density-Functional theory in the form of unrestricted open-shell presentation. Both computational schemes enable to fix the electron correlation, while only the Hartree-Fock theory suggests a set of quantities to be calculated that can quantitatively characterize the electron correlation and be used for a quantitative description of such graphene properties as magnetism, chemical reactivity, and mechanical response. The paper presents concepts and algorithms of the unrestricted Hartree-Fock theory applied for the consideration of magnetic properties of nanographenes, their chemical modification by the example of stepwise hydrogenation, as well as a possible governing the electron correlation by the carbon skeleton deformation.

## 1. Introduction

According to Wikipedia, 'Graphene is an allotrope of carbon, whose structure is one-atom-thick planar sheets of $sp^2$-bonded carbon atoms that are densely packed in a honeycomb crystal lattice' [1]. This definition clearly exhibits a molecular-crystal duality of this extraordinary substance. From the molecular viewpoint, the extraordinariness is provided with the availability of odd electrons that are responsible for the $sp^2$ configuration of valence electrons of carbon atoms. 2D-dimensionality, on the other hand, dictates properties of the regularly packed honeycomb pattern. As a whole, graphene has much in common with both polycondensated benzenoid molecules and 2D-dimensional crystals indeed. Besides, fundamental characteristics of the two forms are tightly interconnected. Thus, as will be discussed below, such seemingly solid state properties as magnetism and mechanics of graphene are of molecular origin.

This peculiar duality is embodied in the computational strategy of graphene, as well. On one hand, the solid state microscopic theory of quasiparticles in 2D space forms the ground for the description of graphene crystal. On the other hand, quantum molecular theory creates the image of the graphene molecule. However, as earlier, the two seemingly different concepts are tightly interconnected at computational level. Thus, the solid state quasiparticles are usually described in the approach based on a unit cell and/or supercell followed with periodic boundary conditions; besides, the unit cell is described at the molecular theory level.

Therefore, the latter lays the foundation of both approaches, whilst rather differently. It is connected with the different origin of the molecular object under study. In the case of solid state approach, the cell and/or supercell should be chosen as a known crystalline motive. When graphene is considered as a molecule, no structural restrictions are introduced in advance. In both cases, it becomes necessary to understand which molecular theory is applicable for the graphene to be studied in the best way.

The current paper concerns peculiarities of the molecular theory of graphene. The latter are obviously connected with both the odd-electron origin of the graphene electron system and these electrons correlation that turns out to play the governing role. The paper is organized as follows. The odd electron correlation is considered in Section 2. Shown based on three criteria is that the odd electrons in graphene are rather strongly correlated. The determination of pure-spin singlet energy is discussed in Section 3. The correlation effect on the magnetic behaviour of graphene is considered in Section 4. Computational strategy of the chemical modification of graphene is presented in Section 5. Section 6 is devoted to the graphene chemistry that due to the electron correlation can be formulated and considered at a quantitative level. The interconnection between the electron correlation and deformation of graphene is discussed in Section 7. Conclusion summarizes all the essences.

## 2. Odd electrons correlation

In spite of formally monatomic crystalline structure of graphene, its properties are evidently governed by the behaviour of odd electrons of hexagonal benzenoid units. The only thing that we know about the behaviour for sure is that the interaction between odd electrons is weak; nevertheless, how weak is it? Is it, nevertheless, enough to provide a tight covalent pairing when two electrons with different spins occupy the same place in space or, oppositely, is it too weak for this and the two electrons are located in different spaces thus becoming spin correlated? This supremely influential molecular aspect of graphene can be visualised on the platform of molecular quantum theory.

When speaking about electron correlation, one must address the problem to the configuration interaction (CI). However, neither full CI nor any its truncated versions, clear and transparent conceptually, can be applied for computations, valuable for graphene nanoscience that requires a vast number of computations to be performed as well as a great number of atoms to be considered [2, 3]. Owing to this, techniques based on single unrestricted open-shell determinants becomes the only alternative. Unrestricted Hartree-Fock (UHF) and unrestricted DFT (spin polarized, UDFT) approaches form the techniques ground and are both sensitive to the electron correlation, but differently due to different dependence of their algorithms on electron spins [4, 5]. Application of the approaches raises two questions: 1) what are criteria that show the electron correlation in the studied system and 2) how much are the solutions of single-determinant approaches informative for a system of correlated electrons.

Answering the first question, three criteria, which highlight the electron correlation at the single-determinant level of theory, can be suggested. Those concern the following characteristic parameters:

**Criterion 1**:

$$\Delta E^{RU} \geq 0 ,$$

where

$$\Delta E^{RU} = E^R - E^U \quad (1)$$

presents a misalignment of energy . Here $E^R$ and $E^U$ are total energies calculated by using restricted and unrestricted versions of the program in use.

**Criterion 2**:

$N_D \neq 0$,

where $N_D$ is the total number of effectively unpaired electrons and is determined as

$$N_D = tr D(r|r') \neq 0 \quad \text{and} \quad N_D = \sum_A D_A. \quad (2)$$

Here, $D(r|r')$ [6] and $D_A$ [7] present the total and atom-fractioned spin density caused by the spin asymmetry due to the location of electrons with different spins in different spaces.

**Criterion 3**:

$\Delta \hat{S}^2 \geq 0$,

where

$$\Delta \hat{S}^2 = \hat{S}_U^2 - S(S+1) \quad (3)$$

presents the misalignment of squared spin. Here, $\hat{S}_U^2$ is the squared spin calculated within the applied unrestricted technique while $S(S+1)$ presents the exact value of $\hat{S}^2$.

Criterion 1 follows from a well known fact that the electron correlation, if available, lowers the total energy [8]. Criterion 2 highlights the fact that the electron correlation is accompanied with the appearance of effectively unpaired electrons that provide the molecule radicalization [6, 7, 9]. These electrons total number depends on interatomic distance: when the latter is under a critical value $R_{cov}^{crit}$, two adjacent electrons are covalently bound and $N_D = 0$. However, when the distance exceeds $R_{cov}^{crit}$ the two electrons become unpaired, $N_D \geq 0$, the more, the larger is the interatomic spacing. In the case of the sp$^2$ C-C bonds, $R_{cov}^{crit} = 1.395$Å [10]. Criterion 3 is the manifestation of the spin contamination of unrestricted single-determinant solutions [7, 9]; the stronger electron correlation, the bigger spin contamination of the studied spin state.

Table 1 presents sets of the three parameters evaluated for a number of right-angled $(n_a,n_z)$ fragments of graphene ($n_a$ and $n_z$ count the numbers of benzenoid units along armchair and zigzag edges of the fragment, respectively [11]), nanographenes NGrs below, by using AM1 version of semiempirical UHF approach implemented in the CLUSTER-Z1 codes [12]. To our knowledge, only these codes allow for computing all the above three parameters simultaneously. As seen in the table, the parameters are certainly not zero, besides greatly depending on the fragment size. The attention should be called to rather large $N_D$ values, both absolute and relative. The finding evidences that the length of C-C bonds in the considered fragments exceed the critical value $R_{cov}^{crit}=1.395$Å. It should be added as well that the relation $N_D = 2\Delta \hat{S}_U^2$, which is characteristic for spin contaminated solutions in the singlet state [7], is rigidly kept over all the fragments.

The data convincingly evidence that the electron correlation in graphene is significant. In view of the correlation, one can answer the second question put above suggesting quantitative explanation of peculiarities of the graphene magnetism, chemistry, and mechanics, much as this has been done for fullerenes [3].

### 3. Pure-spin singlet energy

The wave functions of the unrestricted single-determinant solutions satisfy the operator equations for energy and $z$-projection of spin $S_z$ but do not satisfy the operator equation for squared spin $\hat{S}^2$. This causes a spin contamination of the solution whose extent is determined by $\Delta\hat{S}^2$ from (3). Owing to this, one faces the problem of the evaluation of the energies of pure spin states.

The unrestricted broken symmetry (UBS) approach suggested by Noodleman [13] can be considered as the best way to solve the problem. It is the most widely known among the unrestricted single-determinant computational schemes used in practice, both UHF and UDFT. The UBS approach provides the determination of the exact energy of pure-spin states on the basis of the obtained single-determinant results within each of the computational schemes at the level of theory that is equivalent to the explicit CI. According to approach, the energy of pure-spin singlet state is expressed as

$$E^{PS}(0) = E^U(0) + S_{max} J, \qquad (4)$$

where $E^U(0)$ is the energy of the singlet state of the USB solution while $S_{max}$ is the highest spin multiplicity of the studied odd electron system and $J$ presents the exchange integral, or the magnetic coupling constant [14]

$$J = \frac{E^U(0) - E^U(S_{max})}{S_{max}^2}. \qquad (5)$$

Here, $E^U(S_{max})$ is the energy of the highest-spin-multiplicity state and corresponds to the $S_{max}$-pure-spin state.

Table 2 presents sets of three energies, namely: $E^R(0)$, $E^U(0)$, and $E^{PS}(0)$, alongside with the exchange integrals $J$ related to NGrs considered earlier. As seen in the table, the odd electron correlation causes lowering of not only $E^R(0)$ energy, but $E^U(0)$ as well, whilst much less pronounced in the latter case. It is particularly evident when looking at not absolute energy misalignment $\Delta E^{RPS} = E^R(0) - E^{PS}(0)$ and $\Delta E^{UPS} = E^U(0) - E^{PS}(0)$, but at their percentage quantities $\delta E^{RPS}$ and $\delta E^{UPS}$. As seen from the table, if $\delta E^{RPS}$ deviates from ~20 to 25%, $\delta E^{UPS}$ changes much less within ~2-5%. These values clearly show the measure of incorrectness that is introduced when the graphene sample energy is described by either restricted or unrestricted computational schemes.

Tables 1 and 2 collect data related to NGrs of different size. Addressing this parameter, the presented data can be divided into three groups characterizing by different response to the size expansion. The first group involves $\Delta E^{RU}$, $N_D$, and $\Delta \hat{S}_U^2$ from Table 1 and $\Delta E^{RPS}$, $\Delta E^{UPS}$ from Table 2. All the quantities gradually increase when the size grows due to increasing the number of atoms involved in the pristine structures. The relative parameters $\delta E^{RU}$ and $\delta N_D$ of Table 1 and $\delta E^{RPS}$ of Table 2 form the second group. They all only slightly depend on the NGr size just exhibiting a steady growth of parameters of group 1. Such a behavior is typical for trivial size-dependent reactions on the size growth, is easily foreseen, and is not directly connected with the electron correlation. In contrast to this, quantities $\delta E^{UPS}$ and $J$ of Table 2, which form the third group, demonstrate quite different behavior. Once expected slightly dependent on the species size, similarly to, say, $\delta E^{RPS}$, they both show a significant dependence, gradually decreasing by the absolute value when the size grows. This dependence can be obviously interpreted as the indication of strengthening the electron correlation thus exhibiting the collective character of the event. The finding is expected to lay the foundation of peculiar size-dependent effects for properties that are governed by these parameters, first of which can be addressed to magnetism.

**4. Magnetism of correlated graphene**

Obviously, the odd electrons correlation is a necessary reason for the graphene magnetization. However, this is not enough, since there are requirements concerning the magnetic constant value. Graphene is among singlet bodies, whose magnetic phenomenon may occur as a consequence of mixing the ground singlet state with those of high-spin multiplicity [15] following, say, to the van Fleck mixing promoted by applied magnetic field [16]. Since the effect appears in the first-order perturbation theory, it depends on the $J$ value that determines the energy differences in denominators. Consequently, $J$ should be small by the absolute value to provide noticeable magnetization. Estimated for molecular magnets [17], the phenomenon can be fixed at $|J|$ of $10^{-2}$ -$10^{-3}$ kcal/mol or fewer.

A joint unit cell of graphene involves two atoms that form one C-C bond of the benzenoid unit. Estimation of $J$ value for ethylene and benzene molecule with stretched C-C bonds up to 1.42Å in length gives -13 kcal/mol and –16 kcal/mol, respectively. In spite of ethylene and benzene molecules do not reproduce the unit cell of graphene crystal exactly; a similar $J$ value of the cell constant is undoubted. Owing to this, magnetization of the graphene crystal cannot be observed so that the crystal should demonstrate the diamagnetic behaviour only. The latter is supported both theoretically [18] and empirically (see [19] and references therein). To provide a remarkable magnetization means to drastically decrease the magnetic constant $|J|$, which, in its turn, determines a severe strengthening of the odd electron correlation. Since it is impossible to a regular crystal, let us look what can be expected at the molecular level.

Analyzing data published earlier [20, 21] and addressing the discussion presented in the previous section, one may suggest the NGr size as a regulating factor of the electron correlation. As shown in Table 2, the magnetic constant $|J|$ decreases when NGr becomes larger. When speaking about mixing the ground singlet state with those of high-spin ones, obviously, singlet-triplet mixing is the most influent. As follows from Table 2, the energy gap to the nearest triplet state, equal $2|J|$, for the studied NGrs constitutes 2.8-0.6 kcal/mol. The value is still large to provide a recordable magnetization of these molecular magnets, but the tendency is quite optimistic.

In view of this idea, let us estimate how large should be 'graphene molecule' to provide a noticeable magnetization. As mentioned earlier [21], molecular magnetism can be fixed at $|J|$ of $10^{-2}$ -$10^{-3}$ kcal/mol or fewer. Basing on the data presented in Table 2 and supposing the quantity to be rouph-inversely proportional to the number of odd electrons, we get $N \sim 10^5$. For rectangular NGrs with $N$ odd electrons, the number of carbon atoms constitutes $N = N - 2(n_a + n_z + 1)$ that, according to [11], is determined as

$$N = 2(n_\alpha n_z + n_\alpha + n_z).  \qquad (6)$$

To fit the needed N value, the indices $n_\alpha$ and $n_z$ should be of hundreds, which leads to linear sizes of the NGrs from a few units to tens *nm*. The estimation is rather approximate, but it, nevertheless, correlates well with experimental observations of the magnetization of activated carbon fibers consisting of nanographite domains of ~2 nm in size [22, 23]. Recently, has been reported a direct observation of size-dependent large-magnitude room-temperature ferromagnetism of graphene interpore regions [24, 25]. The maximum effect was observed at the region width of 20 *nm* after which the signal gradually decreased when the width increased. The behaviour is similar to that obtained for fullerene oligomers [26], which led to the suggestion of a scaly mechanism of nanostructured solid state magnetism of the polymerized fullerene $C_{60}$ and was confirmed experimentally.

The obtained results highlight another noteworthy aspect of the graphene magnetism attributing the phenomenon to size ones. The latter means that the graphene magnetization is observed for nanosize samples only, moreover, for samples whose linear dimensions fit a definite interval, while the phenomenon does not take place at either smaller or bigger samples outside of the critical region. An individual benzenoid unit (including benzene molecule) is non-magnetic (only slightly diamagnetic [27]). When the units are joined to form a graphene-like benzenoid cluster, effectively unpaired electrons appear due to weakening the interaction between odd electrons followed by their correlation. The correlation accelerates when the cluster size increases, which is followed with the magnetic constant $|J|$ decreasing until the latter achieves a critical level that provides a noticeable mixing of the singlet ground state with high-spin states for the cluster magnetization to be fixed. Until the enlargement of the cluster size does not violate a molecular (cluster-like) behavior of odd electrons, the sample magnetization will grow. However, as soon as the electron behavior becomes spatially quantized, the molecular character of the magnetization will be broken and will be substituted by that one determined by the electron properties of a unit cell [15]. The critical cluster size is determined by the electron mean free path $l_{el}$. Evidently, when the cluster size exceeds $l_{el}$ the spatial quantization quenches the cluster magnetization. An accurate determination of $l_{el}$ for odd electrons in graphene is not known, but the analysis of a standard data base for electron mean free paths in solids [28] shows the quantity should be ~ 10 *nm*, which is supported by experimental data of 3-7 *nm* electron free path in thin films of Cu-phthalocyanine [29].

Another scenario of getting magnetic graphene is connected with introducing impurity and structural defects in the graphene body. The best illustration of such scenario reality can be found in the latest publication of the Geim team [19] where a paramagnetic behaviour of graphene laminates consisting of 10-50 nm sheets has been recorded after either their fluorination or bombarding by electrons. The treatment provides 'spin-half paramagnetism in graphene induced by point defects'. In both cases, the magnetization is weak and is characterized by one moment per approximately 1,000 carbon atoms, which is explained by the authors by clustering of adatoms and, for the case of vacancies, by the loss of graphene's structural stability. However, the ratio 'one spin per 1,000 carbon atoms' indicates, that, actually, the after-treatment magnetic crystal structure differs from the pristine one and the difference concerns the unit cell that becomes ~33/2 times larger than the previous one. Besides, the unit cell contains one additional spin thus lifting the spin multiplicity to doublet. The latter explains the paramagnetic behaviour of the sample while the size of the cell provides small value of the magnetic constant $|J|$ due to large (~40 nm) cell dimension. Therefore, introduced adatoms and point defects cause a magnetic nanostructuring of the pristine crystal that favors the realization of size-dependent magnetism.

## 5. Computational strategy of the chemical modification of graphene

In view of the correlation, peculiarities of the graphene chemistry can be exhibited at the quantitative level, much as this has been done for fullerenes [3]. As was shown by Takatsuka, Fueno, and Yamaguchi [6], the correlation of weakly interacting electrons is manifested through a density matrix, named as the distribution of 'odd' electrons,

$$D(r|r') = 2\rho(r|r') - \int \rho(r|r'')\rho(r''|r')dr''. \tag{7}$$

The function $D(r|r')$ was proven to be a suitable tool to describe the spatial separation of electrons with opposite spins and its trace

$$N_D = trD(r|r') \tag{8}$$

was interpreted as the total number of these electrons [7, 30]. The authors suggested $N_D$ to manifest the radical character of the species under investigation. Over twenty years later, Staroverov and Davidson changed the term by the 'distribution of *effectively unpaired electrons*' [7, 31] emphasizing that not all odd electrons may be taken off the covalent bonding. Even Takatsuka et al. mentioned [6] that the function $D(r|r')$ can be subjected to the population analysis within the framework of the Mulliken partitioning scheme. In the case of a single Slater determinant, Eq. 8 takes the form [7]

$$N_D = trDS, \tag{9}$$

where

$$DS = 2PS - (PS)^2. \tag{10}$$

Here, $D$ is the spin density matrix $D = P^\alpha - P^\beta$ while $P = P^\alpha + P^\beta$ is a standard density matrix in the atomic orbital basis, and $S$ is the orbital overlap matrix ($\alpha$ and $\beta$ mark different spins). The population of effectively unpaired electrons on atom $A$ is obtained by partitioning the diagonal of the matrix $DS$ as

$$D_A = \sum_{\mu \in A} (DS)_{\mu\mu}, \tag{11}$$

so that

$$N_D = \sum_A D_A. \tag{12}$$

Staroverov and Davidson showed [7] that the atomic population $D_A$ is close to the Mayer free valence index [32] $F_A$ in general case, while in the singlet state $D_A$ and $F_A$ are identical. Thus, a plot of $D_A$ over atoms gives a visual picture of the actual radical electrons distribution [7], which, in its turn, exhibits atoms with enhanced chemical reactivity.

The effectively unpaired electron population is definitely connected with the spin contamination of the UBS solution state. In the case of UBS HF scheme there is a straight relation between $N_D$ and squared spin $\langle S^2 \rangle$ [7]

$$N_D = 2\left( \langle S^2 \rangle - \frac{(N^\alpha - N^\beta)^2}{4} \right), \tag{13}$$

where

$$\langle S^2 \rangle = \left( \frac{(N^\alpha - N^\beta)^2}{4} \right) + \frac{N^\alpha + N^\beta}{2} - \sum_i^{N^\alpha} \sum_j^{N^\beta} \left| \langle \phi_i | \phi_j \rangle \right|^2. \tag{14}$$

Here, $\phi_i$ and $\phi_j$ are atomic orbitals; $N^\alpha$ and $N^\beta$ are the numbers of electrons with spin α and β, respectively.

If UBS HF computations are realized in the *NDDO* approximation (the basis for AM1/PM3 semiempirical techniques) [33], a zero overlap of orbitals leads to $S = I$ in Eq. 10, where $I$ is the identity matrix. The spin density matrix $D$ assumes the form

$$D = (P^\alpha - P^\beta)^2. \tag{15}$$

The elements of the density matrices $P_{ij}^{\alpha(\beta)}$ can be written in terms of eigenvectors of the UHF solution $C_{ik}$

$$P_{ij}^{\alpha(\beta)} = \sum_{k}^{N^{\alpha(\beta)}} C_{ik}^{\alpha(\beta)} * C_{jk}^{\alpha(\beta)} . \tag{16}$$

Expression for $\langle \hat{S}^2 \rangle$ has the form [34]

$$\langle \hat{S}^2 \rangle = \left( \frac{(N^\alpha - N^\beta)^2}{4} \right) + \frac{N^\alpha + N^\beta}{2} - \sum_{i,j=1}^{NORBS} P_{ij}^\alpha P_{ij}^\beta . \tag{17}$$

Within the framework of the *NDDO* approach, the HF-based total $N_D$ and atomic $N_{DA}$ populations of effectively unpaired electrons take the form [35]

$$N_D = \sum_A N_{DA} = \sum_{i,j=1}^{NORBS} D_{ij} \tag{18}$$

and

$$N_{DA} = \sum_{i \in A} \sum_{B=1}^{NAT} \sum_{j \in B} D_{ij} . \tag{19}$$

Here, $D_{ij}$ are elements of spin density matrix $D$ that presents a measure of the electron correlation [6, 7, 36].

Explicit expressions (18) and (19) are the consequence of the wave-function-based character of the UBS HF. Since the corresponding coordinate wave functions are subordinated to definite permutation symmetry, each value of spin $S$ corresponds to a definite expectation value of energy [5]. Oppositely, the electron density ρ is invariant to the permutation symmetry. The latter causes a serious spin multiplicity problem for the UBS DFT [4, 5]. Additionally, the spin density $D(r|r')$ of the UBS DFT depends on spin-dependent exchange and correlation functionals only and cannot be expressed analytically [5]. Since the exchange-correlation composition deviates from one method to the other, the spin density is not fixed and deviates alongside with the composition. To present the real extent of the problem, is enough to mention a comparison of UBS HF and UBS DFT results with those obtained by using computational schemes based on both a complete active space self consistent field (CASSCF) and a multireference configuration interaction (MRCI). The results concerned the description of diradical character of the Cope rearrangement transition state [31]. When CASSCF, MRCI, and UBS HF calculations gave $N_D$ of 1.05, 1.55, and 1.45 *e*, respectively, UBS DFT calculations gave $N_D = 0$. Therefore, experimentally recognized radical character of the transition state was well supported by the former three techniques with quite a small deviation in numerical quantities while UBS DFT results just rejected the radical character of the state. Serious UBS DFT problems are known as well in the relevance to $\langle \hat{S}^2 \rangle$ calculations [37, 38]. These obvious shortcomings make the UDFT approach practically inapplicable in the case when the correlation of weakly interacting electrons is significant. Certain optimism is connected with a particular view on the structure of the density matrix of effectively unpaired electrons developed by the Spanish-Argentine group [9, 36, 39] from one hand and new facilities offered by Jamagouchi's approximately spin-projected geometry optimization method intensely developed by a Japanese team [40, 41], from the other. By sure, this will give a possibility to describe the electron correlation at the density theory level more thoroughly.

In the singlet state, the $N_{DA}$ values are identical to the atom free valences [7] and thus exhibit the atomic chemical susceptibility (ACS) [42]. The $N_{DA}$ distribution over atoms plots a 'chemical portrait' of the studied molecule, whose analysis allows for making a definite choice of the target atom with the highest $N_{DA}$ value be subordinated to chemical attack by an external addend. A typical chemical portrait of graphene fragment highlights edge atoms as those with the highest chemical activities, besides rather irregular, while exhibiting additionally the basal atoms ACS comparable with that one of fullerene $C_{60}$ [17, 18].

This circumstance is the main consequence of the odd electron correlation in graphene in regards its chemical modification. Ignoring the correlation has resulted in a common conclusion about chemical inertness of the graphene atoms with the only exclusion concerning edge atoms. Owing to this, a computationist does not know the place of the first as well as consequent chemical attacks to be possible on the basal plane and has to perform calculations sorting them out over the atoms by using the lowest-total-energy (LTE) criterion (see, for example, [43]). In contrast, basing of the $N_{DA}$ value as a quantitative pointer of the target atom at any step of the chemical attack, one can suggest the algorithmic 'computational synthesis' of the molecule derivatives [44]. In what follows the algorithm-in-action will be illustrated by the example of the hydrogenation of (5, 5) NGr.

## 6. Algorithmic computational design of graphene polyhydride (CH)$_n$

The equilibrium structure of (5,5) NGr alongside with its ACS map, which presents the distribution of atomically-matched effectively unpaired electrons $N_{DA}$ over the fragment atoms, is shown in Fig.1. Panel *b* exhibits the ACS distribution attributed to the atoms positions thus presenting the 'chemical portrait' of the fragment. Different ACS values are plotted in different colouring according to the attached scale. The absolute ACS values are shown in panel *c* according to the atom numbering in the output file. As seen in the figure, 22 edge atoms involving 2x5 *zg* and 2x6 *ach* ones have the highest ACS thus marking the perimeter as the most active chemical space of the fragment. These atoms are highlighted by the fact that each of them posses two odd electrons, the interaction between which is obviously weaker than that for the basal atoms. Providing the latter, the electron correlation and the extent of the electron unpairing are the highest for these atoms, besides bigger for *zg* edges than for *ach* ones.

The hydrogenation of the fragment will start on atom 14 (star-marked in Fig.1c) according to the highest ACS in the output file. The next step of the reaction involves the atom from the edge set as well, and this is continuing until all the edge atoms are saturated by a pair of hydrogen atoms each since all 44 steps are accompanied with the high-rank ACS list where edge atoms take the first place. Thus obtained hydrogen-framed graphene molecule is shown in Fig. 2 alongside with the corresponding ACS map. Two equilibrium structures are presented. The structure in panel *a* corresponds to the optimization of the molecule structure without any restriction. In the second case, positions of edge carbon atoms and framing hydrogen atoms were fixed and the optimization procedure results in the structure shown in panel *c*. In what follows, we shall refer to the two structures as free standing and fixed membranes, respectively. Blue atoms in Fig. 2c alongside with framing hydrogens are excluded from the forthcoming optimization under all steps of the further hydrogenation.

Chemical portraits of the structures shown in Fig. 2b and Fig. 2d are quite similar and reveal the transformation of brightly shining edge atoms in Fig. 1b into dark spots. The addition of two hydrogen atoms to each of the edge ones saturates the valence of the latter completely, which results in zeroing ACS values, as is clearly seen in Fig. 2e. The chemical activity is shifted to the neighboring inner atoms and retains higher in the vicinity of *zg* edges, however, differently in the two cases. The difference is caused by the redistribution of C-C bond lengths of free standing membrane when it is fixed over perimeter, thus providing different starting conditions for the hydrogenation of the two membranes.

Besides the two types of initial membranes, the hydrogenation will obviously depend on other factors, such as 1) the hydrogen species in use and 2) the accessibility of the membranes sides to the hydrogen. Even these circumstances evidence the hydrogenation of graphene to be a complicated chemical event that strongly depends on the initial conditions, once divided into 8 adsorption modes in regards atomic or molecular adsorption; one- or two-side accessibility of membranes; and free or fixed state of the membranes perimeter. Only two of the latter correspond to the experimental observation of hydrogenated specimens discussed in [46], namely: two-side and one-side atomic hydrogen adsorption on the fixed membrane. Only these two modes will be described in details in what follows.

**6.1. Two-side atomic adsorption of hydrogen on fixed membrane**

The hydrogenation concerns the basal plane of the fixed hydrogen-framed membrane shown in Fig. 2c that is accessible to hydrogen atoms from both sides. To facilitate the further presentation of the equilibrium structures, framing hydrogen atoms will not be shown. As seen in Fig. 2e, the first hydrogenation step should occur on basal atom 13 marked by a star. Since the membrane is accessible to hydrogen from both sides, one has to check which deposition of the hydrogen atom, namely, above the carbon plane ('up') or below it ('down') satisfies the LTE criterion. As seen from Chart 1, the up position is somewhat preferential, and the obtained equilibrium structure H1 is shown in Fig. 3. The atomic structure is accompanied with the ACS map that makes it possible to trace the transformation of the chemical activity of the NGr caused by the first deposition. Rows HK (N) in Chart 1 display intermediate graphene hydrides involving K hydrogen atoms adsorbed on the basal plane while H0 is related to (5, 5) NGr with 44 framing hydrogen atoms. N points the number of basal carbon atom in the output file (see Fig. 2e) to which the $K^{th}$ hydrogen atom is attached. Columns $N_{DA}$ and $N_{at}$ present the high rank ACS values and the number of atoms to which they belong. The calculated heat of formation ΔH is subjected to the LTE criterion for selecting the best isomorphs that are shown in the chart by light blue shading. The obtained products form hydrides family 1.

After deposition of hydrogen 1 on basal atom 13, the ACS map has revealed carbon atom 46 for the next deposition (see H1 ACS map in Fig. 3). The LTE criterion favours the down position for the second hydrogen on this atom so that we obtain structure H2 shown in Fig. 3. The second atom deposition highlights next targeting carbon atom 3 (see ACS map of H2 hydride), the third adsorbed hydrogen atom activates target atom 60, the fourth does the same for atom 17, and so forth. Checking up and down depositions in view of the LTE criterion, a choice of the best configuration can be performed and the corresponding equilibrium structures for a selected set of hydrides from H1 to H11 are shown in Fig. 3. As follows from the results, the first 8 hydrogen atoms are deposited on substrate atoms characterized by the largest ACS peaks in Fig. 2b. After saturation of these most active sites, the hydrogen adsorption starts to fill the inner part of the basal plane in a rather non-regular way therewith.

The first hexagon unit with the cyclohexane chairlike motive (cyclohexanoid chair) is formed when the number of hydrogen adsorbates achieves 38. This finding well correlates with experimental observation of a disordered, seemingly occasionally distributed, adsorbed hydrogen atoms on the graphene membrane at similar covering [47]. A complete computational procedure is discussed in details elsewhere [45].

The structure obtained at the end of the 44$^{th}$ step is shown in Fig. 4a. It is perfectly regular, including framing hydrogen atoms thus presenting a computationally synthesized 100% chairlike (5, 5) NGr hydride that is in full accordance with the experimental observation of the graphane crystalline structure [46].

**6.2. One-side atomic adsorption of hydrogen on fixed membrane**

Coming back to the first step of the hydrogenation, which was considered in the previous Section, let us proceed further with the second and all the next steps of the up deposition only. As previously, the choice of the target atom at each step is governed by high-rank $N_{DA}$ values. Figure 5 present a sketch of equilibrium adsorption configurations. As seen in the figure, the sequence of target atoms repeats that one related to the two-side adsorption up to the 10$^{th}$ (actually, to the 11$^{th}$) step, after which the order of target atoms differs from that of the previous case. Starting from the 24$^{th}$ step, a part of the carbon skeleton becomes concave and transforms into a canopy by the final 44$^{th}$ step (see Fig. 4b). However, after a successful adsorption of the 43$^{rd}$ atom, the deposition of the 44$^{th}$ hydrogen not only turns out to be impossible but stimulates desorption of previously adsorbed atom situated in the vicinity of the last target carbon atom. As a consequence, the two hydrogens are coupled, and the hydrogen molecule desorbs from the fragment (see a detailed consideration of such behaviour reasons elsewhere [48]). A peculiar canopy shape of the carbon skeleton of hydride H44 is solely by the formation of the table-like cyclohexanoid units. However, the unit packing is only quasi-regular that may explain the amorphous character of the hydrides formed at the outer surface of graphene ripples observed experimentally [46]. A complete set of the obtained products form hydrides family 2.

As for the hydrogen coverage, Fig. 6 presents the distribution of C-H bond lengths of hydrides H44 of families 1 and 2. In both cases, the distribution consists of two parts, the first of which covers 44 C-H bonds formed at the skeleton edges. Obviously, this part is identical for both hydrides since the bonds are related to framing atoms excluded from the optimization. The second part covers C-H bonds formed by hydrogen atoms attached to the basal plane. As seen in the figure, in the case of hydride 1, C-H bonds are practically identical with the average length of 1.126Å and only slightly deviate from those related to framing atoms. This is just a reflection of the regular graphane-like structure of the hydride shown in Fig.4a similarly to highly symmetric fullerene hydride $C_{60}H_{60}$ [49]. In contrast, C-H bonds on a canopy-like carbon skeleton of hydride 2 are much longer than those in the framing zone, significantly oscillate around the average value of 1.180Å and even achieve value of 1.275Å. In spite of the values greatly exceed a 'standard' C-H bond length of 1.11Å, typical for benzene, those are still among chemical C-H bonds, whilst stretched, since the C-H bond rupture occurs at the C-H distance of 1.72Å [50]. A remarkable stretching of the bonds points to a considerable weakening of the C-H interaction for hydrides 2, which is supported by the energetic characteristics of the hydrides, as well.

**6.3. Energetic characteristics accompanying the nanographene hydrogenation**

Total coupling energy that may characterize the molecule hydrogenation can be presented as

$$E_{cpl}^{tot}(n) = \Delta H_{nHgr} - \Delta H_{gr} - n\Delta H_{at}. \tag{19}$$

Here $\Delta H_{nHgr}$, $\Delta H_{gr}$, and $\Delta H_{at}$ are heats of formation of graphene hydride with $n$ hydrogen atoms, a pristine nanographene, and hydrogen atom, respectively. Since we are mainly interested in the adsorption on basal plane, it is worthwhile to refer the coupling energy related to the basal adsorption to the energy of the framed membrane in the form

$$E_{cpl}^{tot\ bs}(k) = E_{cpl}^{tot}(k+44) - E_{cpl}^{tot\ fr}(44), \tag{20}$$

where $k = n - 44$ (K in Chart 1) numbers hydrogen atoms deposited on the basal plane and $E_{cpl}^{tot\ bs}(k)$ presents the coupling energy counted off the energy of the framed membrane $E_{cpl}^{tot\ fr}(44)$.

The tempo of hydrogenation may be characterized by the coupling energy needed for the addition of each next hydrogen atom. Attributing the energy to the adsorption on the basal plane, the per step energy can be determined as

$$E_{cpl}^{step\ bs}(k) = E_{cpl}^{tot}(k+44) - E_{cpl}^{tot}[(k+44)-1]. \tag{21}$$

Evidently, two main contributions, namely, the deformation of the fragment carbon skeleton (*def*) and the covalent coupling of hydrogen atoms with the substrate resulted in the formation of C-H bonds (*cov*) determine both total and per step coupling energies. Supposing that the relevant contributions can be summed up, one may evaluate them separately. Thus, the total deformation energy can be determined as the difference

$$E_{def}^{tot}(n) = \Delta H_{nHgr}^{sk} - \Delta H_{gr}. \tag{22}$$

Here $\Delta H_{nHgr}^{sk}$ presents the heat of formation of the carbon skeleton of the $n$-th hydride at the $n$-th stage of hydrogenation and $\Delta H_{gr}$ presents the heat of formation of the initial graphene fragment. The value $\Delta H_{nHgr}^{sk}$ can be obtained as a result of one-point-structure determination applied to the $n$-th equilibrium hydride after removing all hydrogen atoms. Attributed to the basal plane, $E_{def}^{tot}(n)$ has the form

$$E_{def}^{tot\ bs}(k) = E_{def}^{tot}(k+44) - E_{def}^{tot\ fr}(44). \tag{23}$$

Here $E_{def}^{tot\ fr}(44)$ presents the deformation energy of the framed membrane.

The deformation energy which accompanies each step of the hydrogenation can be determined as

$$E_{def}^{step\ bs}(k) = \Delta H_{(k+44)Hgr}^{sk} - \Delta H_{[(k+44)-1]Hgr}^{sk}, \qquad (24)$$

where $\Delta H_{(k+44)Hgr}^{sk}$ and $\Delta H_{[(k=44)-1]Hgr}^{sk}$ match heats of formation of the carbon skeletons of the relevant hydrides at two subsequent steps of hydrogenation.

Similarly, the total and per step chemical contributions caused by the formation of C-H bonds on the basal plane can be determined as

$$E_{cov}^{tot\ bs}(k) = E_{cpl}^{tot\ bs}(k) - E_{def}^{tot\ bs}(k) \qquad (25)$$

and

$$E_{cov}^{step\ bs}(k) = E_{cov}^{tot}(k+44) - E_{cov}^{tot}(k+44-1). \qquad 26)$$

Figure 7 displays the calculated total energies for hydrides 1 and hydrides 2. The relevant per step energies are shown in Fig.8. As seen in Fig.7, the total energies $E_{cpl}^{tot\ bs}$ of both hydrides are negative by sign and gradually increase by the absolute value when the number of adsorbed atoms increases. Besides, the absolute value growth related to hydrides 2 is evidently slowing down starting at step 11 in contrast to the continuing growth for hydrides 1. This retardation is characteristic for other two energies presented in Fig.7 thus quantitatively distinguishing hydrides 2 from hydrides 1. The retardation of the growth of both $E_{cpl}^{tot\ bs}$ and $E_{cov}^{tot\ bs}$ energies obviously show that the addition of hydrogen to the fixed membrane of hydrides 2 at coverage higher than 30% is more difficult than in the case of hydrides 1. This conclusion is supported by the behaviour of per step energies $E_{cpl}^{step\ bs}$ and $E_{cov}^{step\ bs}$ plotted in Fig.8a and 8c. If, in the case of hydrides 1, the energy values oscillate around steady average values of -52 kcal/mol and -72 kcal/mol for $E_{cpl}^{step\ bs}$ and $E_{cov}^{step\ bs}$, respectively, in the case of hydrides 2, $E_{cpl}^{step\ bs}$ oscillates around average values that grow from -64 kcal/mol to -8 kcal/mol. Similar $E_{cov}^{step\ bs}$ oscillations occur around a general level that starts at -88 kcal/mod and terminates at -8 kcal/mol. Therefore, the reaction of the chemical attachment of hydrogen atoms for hydrides 1 is thermodynamically profitable through over the covering that reaches 100% limit. In contrast, the large coverage for hydrides 2 becomes less and less profitable so that at final steps adsorption and desorption become competitive thus resulting in desorption of hydrogen molecules, which was described in Section 6.2.

An attention should be given to changing the deformation of the carbon skeleton caused by $sp^2 \rightarrow sp^3$ transformation of the carbon atom electron configuration. Gradually increased by value for both hydride families, the energy $E_{def}^{tot\ bs}$ shown in Fig.7 describes strengthening the deformation in due course of growing coverage of the basal plane. Irregular dependence of $E_{def}^{step\ bs}$ on covering presented in Fig.8b allows for speaking about obvious topochemical character of a multistep attachment of hydrogen atoms to the membrane basal plane. The topochemistry in this case implies chemical reactions occurred in a space subordinated to restricting conditions (see [51] and references therein) like reactions on the solid surfaces [52]. The disclosed dependences may serve as a direct manifestation of the reactions of such kind.

## 7. Mechanics of correlated graphene

Deformation of graphene is tightly connected with odd electron correlation since it concerns changing interatomic distances. The latter is very important regulator of the correlation extent thus increasing it when the distance grows. Obviously, strengthening of electron correlation results in the growths of the number of effectively unpaired electrons $N_D$. A typical example of such behavior can be illustrated by the dependence of $N_D$ on interatomic distance that occurs at homolitic cleavage of C-C bond in ethylene molecule presented in Fig. 9. Stretching the bond from its equilibrium value of 1.326Å up to $R_{crit}$=1.395Å does not cause the appearance of the unpaired electrons. Above $R_{crit}$ the number $N_D$ gradually increases up to a clearly vivid knee that is characterized by $N_D \cong 2$ at $R$=1.76Å, which evidences a complete radicalization of two odd electrons. Further stretching concerns mainly two σ electrons that gradually become unpaired as well resulting in $N_D \cong 4$ at 2.5Å.

    A similar stretching of C-C bonds can be highlighted when comparing the carbon skeletons of the pristine (5, 5) NGr and those canopy-like and basket-like ones subjected to one-side hydrogen adsorption on either fixed or free standing membrane, respectively [48]. Figure 10 presents the views of the skeletons alongside with the distribution of their C-C bond lengths. As seen in the figure, C-C bonds of both deformed skeletons are elongated, whilst the summary elongation for the basket-like skeleton is evidently bigger than that one for the canopy-like one. The elongation is restricted by the bond length 1.53Å, which is dictated by $sp^3$ configuration of carbon atoms due to hydrogenation. Naturally, the accumulated deformation may cause some bonds breaking, which occurs for bond 2 of the basket-like skeleton. As a whole, changes in the C-C bond lengths presented in Fig.10 result in decreasing magnetic constant *J* by the absolute value from -1.43 kcal/mol for the pristine (5, 5) NGr to -0.83 and -0.59 kcal/mol for the canopy-like and basket-like skeletons. Simultaneously, $N_D$ increases from 31*e* to 46*e* and 54*e*, respectively. Both findings evidence an undoubted strengthening of the odd electron correlation caused by the chemically-stimulated deformation of the carbon skeleton.

    Yet another evidence of the deformation effect is presented in Fig.11. The figure shows the redistribution of unpaired electrons density over the skeleton atoms caused by the deformation. As seen in the figure, the skeleton electron-density image greatly changes when the electron correlation becomes stronger (draw attention on the vertical scales of plottings presented in figure). Consequently, if observed by HRTEM, the basket-like skeleton might have look much brighter than the canopy-like one and especially than the least bright pristine NGr. In view of finding, it is naturally to suggest that raised above the substrate and deformed areas of graphene in the form of bubbles, found in a variety of shapes on different substrates [53, 54], reveal peculiar electron-density properties just due to stretching deformation that results in strengthening the odd electron correlation. Small (5, 5) NGr presented in the figure cannot pretend to simulate the picture observed for micron bubbles, but it exhibits the general tendency that might take place in bubbles, as well. In view of the obvious strengthening of the odd electron correlation caused by the deformation, this explanation looks more natural than that proposed from the position of an artificial 'gigantic pseudo-magnetic field' [53].

Abovementioned considerable decreasing of magnetic constants $J$ stimulated by deformation allows for suggesting a peculiar magnetic behaviour of the deformed graphene regions, such as, say, bubbles, stimulated by both their size and curvature. The two parameters obviously favour decreasing in the constant values thus promoting the appearance of magnetic response localized in the bubble regions.

Besides the formation of bubbles caused by ultrastrong adhesion of graphene membranes to different substrates [55], the deformation of graphene can be caused by the application of external stress. Quantum molecular theory suggests considering the graphene deformation and rupture in terms of a mechanochemical reaction [56-58]. The quantum chemical realization of the approach is based on the coordinate-of-reaction concept for the purpose of introducing a mechanochemical internal coordinate (MIC) that specifies a deformational mode. The related force of response is calculated as the energy gradient along the MIC, while the atomic configuration is optimized over all of the other coordinates under the MIC constant-pitch elongation. When applied to the description of the deformation of both (5, 5) nanographene [56, 57] and (5, 5) nanographane [58] under uniaxial tension, the calculations highlighted a pronounced changing in the number of effectively unpaired electrons $N_D$ of the sample in due course of its deformation. As shown, the changing is different when the deformation occurs either along or normal to the chains of C-C bonds. However, in all cases the changing is quite significant pointing to a considerable strengthening of odd electron correlation due to changes in interatomic spacings. A detailed consideration of a possible regulating mission of stress with respect to the enhancement of chemical reactivity of carbon atoms and magnetic behaviour of the loaded sample obviously deserves a further thorough study.

## 8. Discussion and conclusive remarks

Data presented in the current paper have shown that the correlation of odd electrons in graphene is significant so that the species should be attributed to correlated ones. This finding considerably complicates the computational strategy addressing it to multireference computational schemes. Owing to serious problems related to the realization of such schemes in practice, a compromise can be suggested by using single-determinant approaches based on either Hartree-Fock or density-functional theory in the form of unrestricted open-shell presentation. Both computational schemes can fix the electron correlation, while only the Hartree-Fock theory suggests a set of quantities to be calculated that can quantitatively characterize the electron correlation and be used for a quantitative description of such graphene properties as magnetism, chemical reactivity, and mechanical response.

Magnetic constant $J$, the inverse value of which directly characterizes the strength of the electron correlation, lays the foundation of the description of magnetic properties of graphene. As shown, the constant is ~-16kcal/mol for a regular crystal, which is too much for the magnetism to be observable. Oppositely to crystalline graphene, nondeformed and nondestructured nanographenes can be magnetic, since the magnetic constant rough-inversely depends on the NGr size. The latter determines the correlation zone and provides a recordable magnetization if is from a few units to tens nanometers. Small enough $J$ values can be provided as well by magnetic nanostructuring of rather large graphene pieces caused by the introduction of half-spin impurities or defects into the graphene body. Additionally, the odd number of the latter may change the singlet spin multiplicity of the ground state. Both scenarios find their verification in practice.

The correlated-electron algorithm-in-action has been applied in the current paper to the chemical modification of (5, 5) NGr by the example of its algorithmic stepwise hydrogenation. As shown, both the hydrogenation of NGr itself and the final hydrides depend on several external factors, namely: 1) the state of the fixation of graphene substrate; 2) the accessibility of the substrate sides to hydrogen; and 3) molecular or atomic composition of the hydrogen vapour. These circumstances make both computational consideration and technology of the graphene hydrogenation multimode with the number of variants not less than eight if only molecular and atomic adsorption does not occur simultaneously. Thus, in full agreement with experimental evidence, a regular chair-like cyclohexanoid structure known as graphane can be obtained only for fixed membranes accessible to hydrogen atoms from both sides. Oppositely, one-side hydrogenation of the membrane results in irregular quasi-amorphous structure. A detailed consideration of all variants should be mandatory included in any serious project aimed at application of the hydrogen-graphene-based nanomaterials in general and for hydrogen-stored fuel cells, in particular.

The electron correlation is highly sensitive to the graphene deformation since the interatomic spacing is the main regulator of the former. Both static and dynamic deformation may influence the action. Static deformation, considered in the paper, is caused by one-side hydrogenation of the pristine sample. When hydrogen atoms are removed, the carbon skeleton keeps its concave, either a canopy-like or basket-like shape supported by stretched C-C bonds. The stretching stimulates both a significant increasing of the number of effectively unpaired electrons and a remarkable decreasing the magnetic constant absolute value. The former changes the electron-density image of the sample and evidence the enhancement of its chemical reactivity. The latter may result in the deformation-stimulated magnetization of the sample. In view of these findings, an alternative, correlated-electron explanation of peculiarities related the density images of the graphene bubbles found on different substrates has been suggested.

The dynamic deformation is illustrated by the example of the (5, 5) NGr uniaxial tension. As shown, the strengthening of the electron correlation accompanies each step of the deformation. This is followed by both enhancement of chemical reactivity and magnetic ability. In spite of predominantly plastic character of the graphene deformation, the latter can be applied to regulate both abilities of the sample.

The odd electron correlation is not a prerogative of graphene only. Similar phenomenon is characteristic for all $sp^2$ nanocarbons, including both fullerenes and nanotubes [3]. The only preference of graphene consists in much larger variety of cases when this inherent characteristic of the class can be visualized.

## ACKNOWLEDGEMENTS

The authors immensely appreciate fruitful discussions with I.L.Kaplan who draw her attention onto problems in DFT with the total spin and degenerate states thus stimulating to think about the computational strategy of $sp^2$ nanocarbons.

## REFERENCES

1. Geim, A. K.; Novoselov, K. S. Nature Mat. 2007, 6, 183.
2. Davidson, E.R.; Clark, A.E. Phys Chem. Chem. Phys. 2007, 9, 1881.
3. Sheka, E.F. Fullerenes: Nanochemistry, Nanomagnetism, Nanomedicine, Nanophotonics. CRC Press, Taylor and Francis Group: Boca Raton. 2011.
4. Davidson, E. Int. J. Quant. Chem. 1998, 69, 214.

5. Kaplan, I. Int. J. Quant. Chem. 2007,107, 2595.
6. Takatsuka, K.; Fueno, T.; Yamaguchi, K. Theor.Chim.Acta 1978, 48, 175.
7. Staroverov, V.N.; Davidson, E.R. Chem.Phys.Lett. 2000, 330, 161.
8. Benard, M. J.Chem.Phys.1979, 71, 2546.
9. Lain, L.; Torre, A.; Alcoba, D.R.; Bochicchio, R.C. Theor. Chem. Acc. 2011, 128, 405.
10. Sheka, E.F.; Chernozatonskii, L.A. J. Phys. Chem. A 2007, 111, 10771.
11. Gao, X.; Zhou, Z.; Zhao, Y.; Nagase, S.; Zhang, S.B.; Chen, Z. J. Phys. Chem. A 2008, 112, 12677.
12. Zayets, V.A. CLUSTER-Z1: Quantum-Chemical Software for Calculations in the s,p-Basis", Institute of Surface Chemistry, Nat. Ac.Sci. of Ukraine: Kiev, 1990.
13. Noodleman, L. J Chem Phys. 1981, 74, 5737.
14. Illas, F.; Moreira, I. de P.R.; de Graaf,C.; Barone, V. Theor. Chem. Acc. 2000,104, 265.
15. Zvezdin, A.K.; Matveev, V.M.; Mukhin, A.A. et al. Redkozemeljnyje iony v magnito-uporjadochennykh kristallakh (Rear Earth Ions in Magnetically Ordered Crystals). Nauka: Moskva, 1985.
16. Van Fleck, J.H. The Theory of Electric and Magnetic Susceptibilities. Oxford. 1932.
17. Kahn, O. Molecular Magnetism. VCH, New York. 1993.
18. Koshino, M.; Ando, T. Phys. Rev. B 2007, 75, 235333.
19. Nair, R. R.; Sepioni, M. ; Tsai, I-L.; Lehtinen, O.; Keinonen, J.; Krasheninnikov, A. V.; Thomson, T.; Geim, A. K.; Grigorieva, I. V. Nature Phys. 2012, doi:10.1038/nphys2183
20. Sheka, E.F.; Chernozatonskii, L.A. Int. J. Quant. Chem. 2010, 110, 1466.
21. Sheka, E.F.; Chernozatonskii, L.A. JETP 2010, 110, 121.
22. Shibayama, Y.; Sato, H.; Enoki, T.; Endo, M. Phys. Rev. Lett. 2000, 84, 1744.
23. Enoki, T.; Kobayashi, Y. J. Mat. Chem. 2005, 15, 3999.
24. Tada, K.; Haruyama, J.; Yang, H. X.; Chshiev, M.; Matsui, T.; Fukuyama, H. Appl. Phys. Lett. 2011, 99, 183111.
25. Tada, K.; Haruyama, J.; Yang, H. X.; Chshiev, M.; Matsui, T.; Fukuyama, H. Phys Rev Lett. 2011, 107, 217203.
26. Sheka, E.F., Zayets, V.A., Ginzburg, I.Ya. JETP 2006, 103, 728.
27. Boeker, G.F. Phys. Rev. 1933, 43, 756.
28. Seach, M.P.; Dench, W.A. Surf. Interf. Analysis 2004, 1, 2.
29. Komolov, S.A.; Lazneva, E.F.; Komolov, A.S. Pis'ma JTF 2003, 29 (23), 13.
30. Takatsuka, K.; Fueno, T. J. Chem. Phys. 1978, 69, 661.
31. Staroverov, V.N.; Davidson, E.R. J. Am. Chem. Soc. 2000, 122, 186.
32. Mayer, I. Int. J. Quant. Chem. 1986, 29, 73.
33. Dewar, M.J.S.; Thiel, W. J. Am. Chem. Soc. 1977, 99, 4899.
34. Zhogolev, D.A.; Volkov, V.B. Methods, Algorithms and Programs for Quantum-Chemical Calculations of Molecules (in Russian). Naukova Dumka: Kiev. 1976.
35. Sheka, E.F.; Zayets, V.A. Russ. Journ. Phys. Chem. 2005, 79, 2009.
36. Lain, L.; Torre, A.; Alcoba, D.R.; Bochicchio, R.C. Chem. Phys. Lett. 2009, 476, 101.
37. Wang, J.; Becke, A.D.; Smith, V.H., Jr. J. Chem. Phys. 1995,102, 3477.
38. Cohen, A.J.; Tozer, D.J.; Handy, N.C. J. Chem. Phys. 2007, 126, 214104.
39. Lobayan, R.M.; Bochicchio,R.C.; Torre, A.; Lain, L. J. Chem. Theory Comp. 2011, 7, 979.
40. Kitagawa, Y.; Saito, T.; Ito, M.; Shoji, M.; Koizumi, K.; Yamanaka, S.; Kawakami, T.; Okumura, M.; Yamaguchi, K. Chem. Phys. Lett. 2007, 442, 445.

41. Kitagawa,Y.; Saito,T.; Nakanishi, Y.; Kataoka, Y.; Matsui, T.; Kawakami, T.; Okumura, M.; Yamaguchi, K. J. Phys. Chem. A 2009, 113, 15041.
42. Sheka, E.F. Int. Journ. Quant. Chem. 2007, 107, 2935.
43. Allouche, A.; Jelea, A.; Marinelli, F.; Ferro, Y. Phys. Scr. 2006, 124, 91.
44. Sheka, E.F. JETP 2010, 111, 395.
45. Sheka, E.F.; Popova, N.A. J. Mod. Mol. 2012, doi: : 10.1007/s00894-012-1356-9.
46. Elias, D.C.; Nair, R.R.; Mohiuddin, T.M.G.; Morozov, S.V.; Blake, P.; Halsall, M.P.; Ferrari, A.C.; Boukhvalov, D.W.; Katsnelson, M.I.; Geim, A.K.; Novoselov, K.S. Science 2009, 323, 610.
47. Meyer, J.C.; Girit, C.O.; Crommie, M.F.; Zettl, A. Nature 2008, 454, 319.
48. Sheka, E.F.; Popova, N.A. arXiv 1201.xxxxv1[cond-mat.mtrl-sci] 2012
49. Sheka, E.F. J. Mod. Mol. 2011, 17, 1973.
50. Sheka, E.F.; Popova, N.A. arXiv:1111.1530v1 [physics.chem-ph] 2011.
51. Sheka, E.F. (in press)
52. Schmidt, G.M.J. Pure Appl. Chem. 1971, **27,** 647.
53. Levy, N.; Burke, S. A.; Meaker, K. L.; Panlasigui, M.; Zettl, A.; Guinea, F.; Neto, A. H. C.; Crommie, M. F. Science 2010, 329, 544.
54. Georgiou, T.; Britnell, L.; Blake, P.; Gorbachev, R. V.; Gholinia, A.; Geim, A. K.; Casiraghi, C.; Novoselov, K. S. Appl. Phys. Lett. 2011, 99, 093103.
55. Koenig, S.P.; Boddeti, N.G.; Dunn, M.L.; Bunch, J.S. Nature Nanotechn. 2011, 6, 543.
56. Sheka, E.F.; Popova, N.A.; Popova, V.A.; Nikitina, E.A.; Shaymardanova, L.K. J. Exp. Theor. Phys. 2011, 112, 602.
57. Sheka, E.F.; Popova, N.A.; Popova, V.A.; Nikitina, E.A.; Shaymardanova, L.K. J. Mol. Mod. 2011, 17, 1121.
58. Sheka, E.F.; Popova, N.A. J. Phys. Chem. C 2011, 115**,** 23745.

**Figure captions**

**Figure 1**. Top and side views of the equilibrium structure of (5,5) nanographene (a) and ACS distribution over atoms in real space (b) and according to atom numbers in the output file (c) [45].

**Figure 2**. Equilibrium structures of free standing (top and side views) (a) and fixed (c) (5,5) graphene membrane and ACS distribution over atoms in real space (b, d) and according to the atom numbers in output file (e). Light gray histogram plots ACS data for the pristine (5,5) nanographene. Curve and black histogram are related to membranes in panels *a* and *c*, respectively [45].

**Figure 3.** Equilibrium structures (left) and real-space ACS maps (right) of hydrides 1 related to initial stage of the basal-plane hydrogenation. HKs denote hydrides with K hydrogen atoms deposited on the membrane [45].

**Figure 4.** Top and side views of the equilibrium structures of hydrides formed at the atomic adsorption of hydrogen on the fixed (5,5) graphene membrane, accessible to the adsorbate from both (a) and one (b) sides.

**Figure 5.** Equilibrium structures of hydrides 2 formed at the one-side basal-plane hydrogenation of the fixed (5,5) graphene membrane. See caption to Fig.3. The desorbed hydrogen molecule from hydride 44 is marked by ellipse (see text) [48].

**Figure 6.** C-H bond length distribution for H44 hydrides of families 1 (1) and 2 (2) [48].

**Figure 7.** Total energies of coupling $E_{cpl}^{tot\ bs}$ (triangles), deformation $E_{def}^{tot\ bs}$ (squares) and covalent bonding $E_{cov}^{tot\ bs}$ (circles) via the number of hydrogen atoms deposited on the basal plane for hydrides 1 (dark gray) and 2 (light blue) according to Exs. (7), (10), and (12), respectively, [48].

**Figure 8.** Per step energies of coupling $E_{cpl}^{step\ bs}$ (a), deformation $E_{def}^{step\ bs}$ (b) and covalent bonding $E_{cov}^{step\ bs}$ (c) via the number of hydrogen atoms deposited on the basal plane for hydrides 1 (dark gray) and 2 (light blue) according to Exs. (8), (11), and (13), respectively, [48].

**Figure 9.** The total number of effectively unpaired electrons $N_D$ of ethylene versus the C-C distance. $R_{cov}^{C-C}$ marks the extreme distance that corresponds to the completion of the covalent bonding. $R_{rad}^{C-C}$ matches a completion of homolytic bond cleavage. Two vertical arrows mark the interval of the C-C bond lengths characteristic for sp$^2$ nanocarbons.

**Figure 10.** C-C bond length distribution for carbon skeletons of the pristine (5, 5) NGr (gray filled region), canopy-like (dark triangles) and basket-like (gray balls) fixed membranes.

**Figure 11.** Effectively-unpaired-electron-density images of the carbon skeletons of the pristine (5, 5) NGr (a), canopy-like (b) and basket-like (c) fixed membranes.

**Chart 1.** Explication of the subsequent steps of the hydrogenation of (5, 5) nanographene. $N_{at}$ number carbon atoms, $N_{DA}$ presents the atomic chemical susceptibility of the atoms, ΔH is the energy of formation of the hydrides in kcal/mol.

| H0 | | H1 (13) | | H2 (46) | | H3 (3) | | H4 (60) | | H5 (17) | | H6 (52) | | H7 (9) | |
|---|---|---|---|---|---|---|---|---|---|---|---|---|---|---|---|
| $N_{at}$ | $N_{DA}$ | $N_{at}$ | $N_{DA}$ | $N_{at}$ | $N_{DA}$ | $N_{at}$ | $N_{DA}$ | $N_{at}$ | $N_{DA}$ | $N_{at}$ | $N_{DA}$ | $N_{at}$ | $N_{DA}$ | $N_{at}$ | $N_{DA}$ |
| 13 | 0,87475 | up | | up | | up | | up | | up | | up | | up | |
| 46 | 0,87034 | 46 | 0,62295 | 3 | 0,56119 | 60 | 0,55935 | 17 | 0,54497 | 52 | 0,49866 | 56 | 0,47308 | 56 | 0,47318 |
| 3 | 0,56633 | 3 | 0,61981 | 60 | 0,55965 | 17 | 0,54302 | 56 | 0,50027 | 56 | 0,49856 | 9 | 0,47072 | 12 | 0,42593 |
| 60 | 0,55862 | 60 | 0,56055 | 17 | 0,54299 | 56 | 0,51849 | 52 | 0,49996 | 9 | 0,47193 | 12 | 0,36783 | 8 | 0,42263 |
| 56 | 0,51946 | ΔH = 126.50 | | ΔH = 101.88 | | ΔH = 96.88 | | ΔH = 99.77 | | ΔH = 95.00 | | ΔH = 96.82 | | ΔH = 97.50 | |
| 17 | 0,50841 | down | | down | | down | | down | | down | | down | | down | |
| 52 | 0,45883 | 46 | 0,62197 | 3 | 0,56174 | 60 | 0,55906 | 17 | 0,54576 | 52 | 0,49684 | 9 | 0,47392 | 56 | 0,46883 |
| 9 | 0,45843 | 3 | 0,62099 | 60 | 0,55926 | 17 | 0,54279 | 56 | 0,49878 | 56 | 0,49602 | 56 | 0,47247 | 12 | 0,41783 |
| 58 | 0,33071 | 60 | 0,55979 | 17 | 0,54297 | 56 | 0,51736 | 52 | 0,49782 | 9 | 0,47141 | 12 | 0,36857 | 8 | 0,41459 |
| 19 | 0,31539 | ΔH = 127.54 | | ΔH = 98.19 | | ΔH = 101.90 | | ΔH = 95.99 | | ΔH = 99.40 | | ΔH = 96.23 | | ΔH = 98.92 | |

| H8 (56) | | H9 (12) | | H10 (57) | | H11 (53) | | H12 (58) | | H13 (33) | | H14 (30) | | H15 (34) | |
|---|---|---|---|---|---|---|---|---|---|---|---|---|---|---|---|
| $N_{at}$ | $N_{DA}$ | $N_{at}$ | $N_{DA}$ | $N_{at}$ | $N_{DA}$ | $N_{at}$ | $N_{DA}$ | $N_{at}$ | $N_{DA}$ | $N_{at}$ | $N_{DA}$ | $N_{at}$ | $N_{DA}$ | $N_{at}$ | $N_{DA}$ |
| up | | up | | up | | up | | up | | up | | up | | up | |
| 12 | 0,42462 | 57 | 0,42824 | 53 | 0,30873 | 58 | 0,54014 | 33 | 0,33339 | 30 | 0,40225 | 34 | 0,34848 | 35 | 0,51203 |
| 8 | 0,42404 | 53 | 0,42131 | 8 | 0,30851 | 30 | 0,32126 | 8 | 0,30396 | 34 | 0,3618 | 31 | 0,33662 | 31 | 0,36904 |
| 53 | 0,41554 | 47 | 0,30877 | 2 | 0,2956 | 8 | 0,32075 | 2 | 0,29194 | 8 | 0,34811 | 8 | 0,2934 | 2 | 0,29541 |
| ΔH = 99.83 | | ΔH = 118.04 | | ΔH = 79.46 | | ΔH = 74.17 | | ΔH = 73.81 | | ΔH = 65.87 | | ΔH = 70.90 | | ΔH = 73.11 | |
| down | | down | | down | | down | | down | | down | | down | | down | |
| 12 | 0,4232 | 57 | 0,42759 | 8 | 0,30801 | 58 | 0,58671 | 33 | 0,32811 | 30 | 0,40175 | 34 | 0,34603 | 35 | 0,5002 |
| 57 | 0,42303 | 53 | 0,4216 | 2 | 0,29368 | 8 | 0,32092 | 8 | 0,30638 | 34 | 0,3625 | 31 | 0,3272 | 31 | 0,36752 |
| 8 | 0,42294 | 47 | 0,31082 | 53 | 0,29305 | 30 | 0,30986 | 2 | 0,28938 | 8 | 0,34298 | 2 | 0,29516 | 2 | 0,29455 |
| ΔH = 98.98 | | ΔH = 89.19 | | ΔH = 107.96 | | ΔH = 101.20 | | ΔH = 68.89 | | ΔH = 84.89 | | ΔH = 67.58 | | ΔH = 71.83 | |

**Table 1**. Identifying parameters of the odd electron correlation in right-angle nanographenes

| Fragment $(n_a, n_z)$ | Odd electrons $N_{odd}$ | $\Delta E^{RU}$ [1] kcal/mol | $\delta E^{RU}$ % [2] | $N_D$, $e^-$ | $\delta N_D$, % [2] | $\Delta \mathcal{S}_U^{2}$ |
|---|---|---|---|---|---|---|
| (5, 5) | 88 | 307 | 17 | 31 | 35 | 15.5 |
| (7, 7) | 150 | 376 | 15 | 52.6 | 35 | 26.3 |
| (9, 9) | 228 | 641 | 19 | 76.2 | 35 | 38.1 |
| (11, 10) | 296 | 760 | 19 | 94.5 | 32 | 47.24 |
| (11, 12) | 346 | 901 | 20 | 107.4 | 31 | 53.7 |
| (15, 12) | 456 | 1038 | 19 | 139 | 31 | 69.5 |

[1] AM1 version of UHF codes of CLUSTER-Z1 [12]. Presented energy values are rounded off to an integer

[2] The percentage values are related to $\delta E^{RU} = \Delta E^{RU} / E^R(0)$ and $\delta N_D = N_D / N_{odd}$, respectively

**Table 2.** Energies of singlet ground state of correlated nanographenes[1], *kcal/mol*

| Fragment $(n_a, n_z)$ | $E^R(0)$ | $E^U(0)$ | $E^{PS}(0)$ | $\Delta E^{RPS}$ | $\delta E^{RPS}$ [2] % | $\Delta E^{UPS}$ | $\delta E^{UPS}$ [2] % | $J$ *kcal/mol* |
|---|---|---|---|---|---|---|---|---|
| (5, 5) | 1902 | 1495 | 1432 | 470 | 24.70 | 63 | 4.39 | -1.429 |
| (7, 7) | 2599 | 2223 | 2156 | 443 | 17.03 | 67 | 3.09 | -0.888 |
| (9, 9) | 3419 | 2778 | 2710 | 709 | 20.75 | 68 | 2.53 | -0.600 |
| (11, 10) | 4072 | 3312 | 3241 | 831 | 20.42 | 71 | 2.20 | -0.483 |
| (11, 12) | 4577 | 3676 | 3606 | 971 | 21.22 | 70 | 1.95 | -0.406 |
| (15, 12) | 5451 | 4413 | 4339 | 1112 | 20.40 | 74 | 1.70 | -0.324 |

[1] AM1 version of UHF codes of CLUSTER-Z1. Presented energy values are rounded off to an integer.
[2] The percentage values are related to $\delta E^{RPS} = \Delta E^{RPS} / E^R(0)$ and $\delta E^{UPS} = \Delta E^{UPS} / E^U(0)$, respectively.

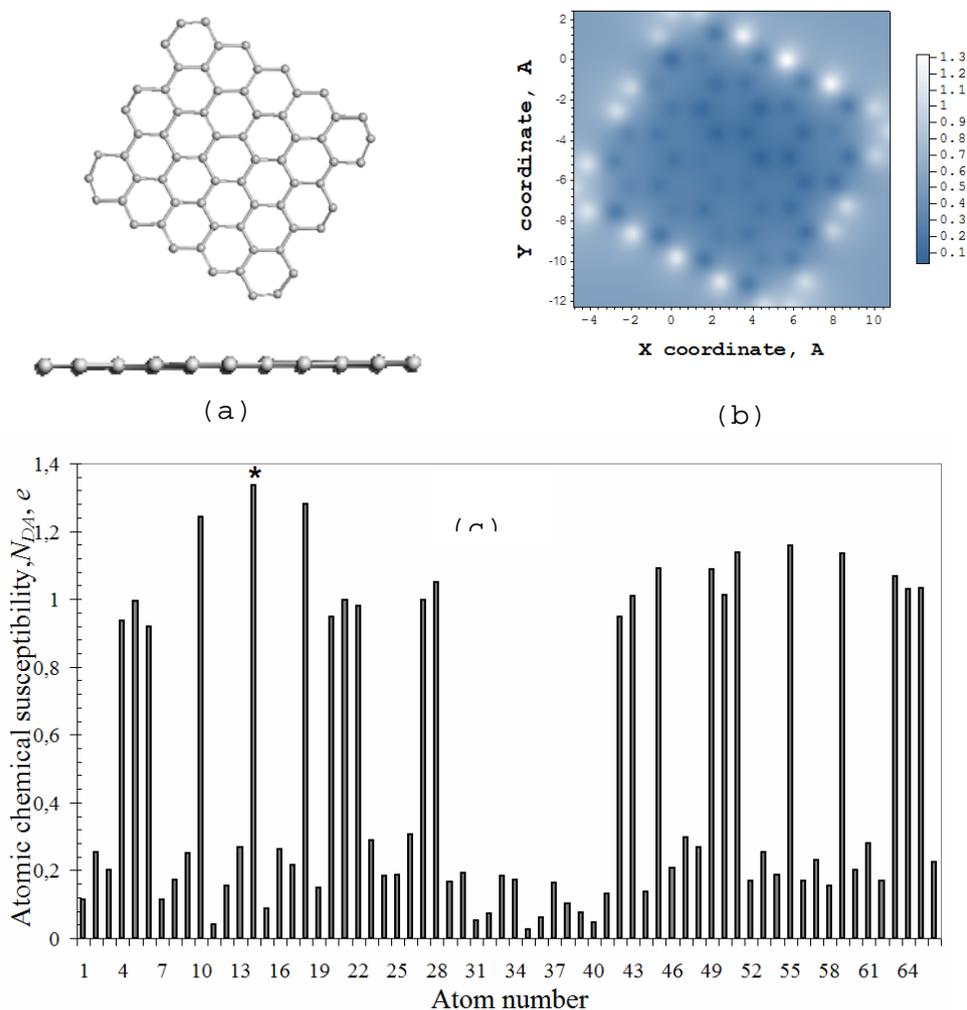

(a)   (b)

(c)

**Figure 1**.

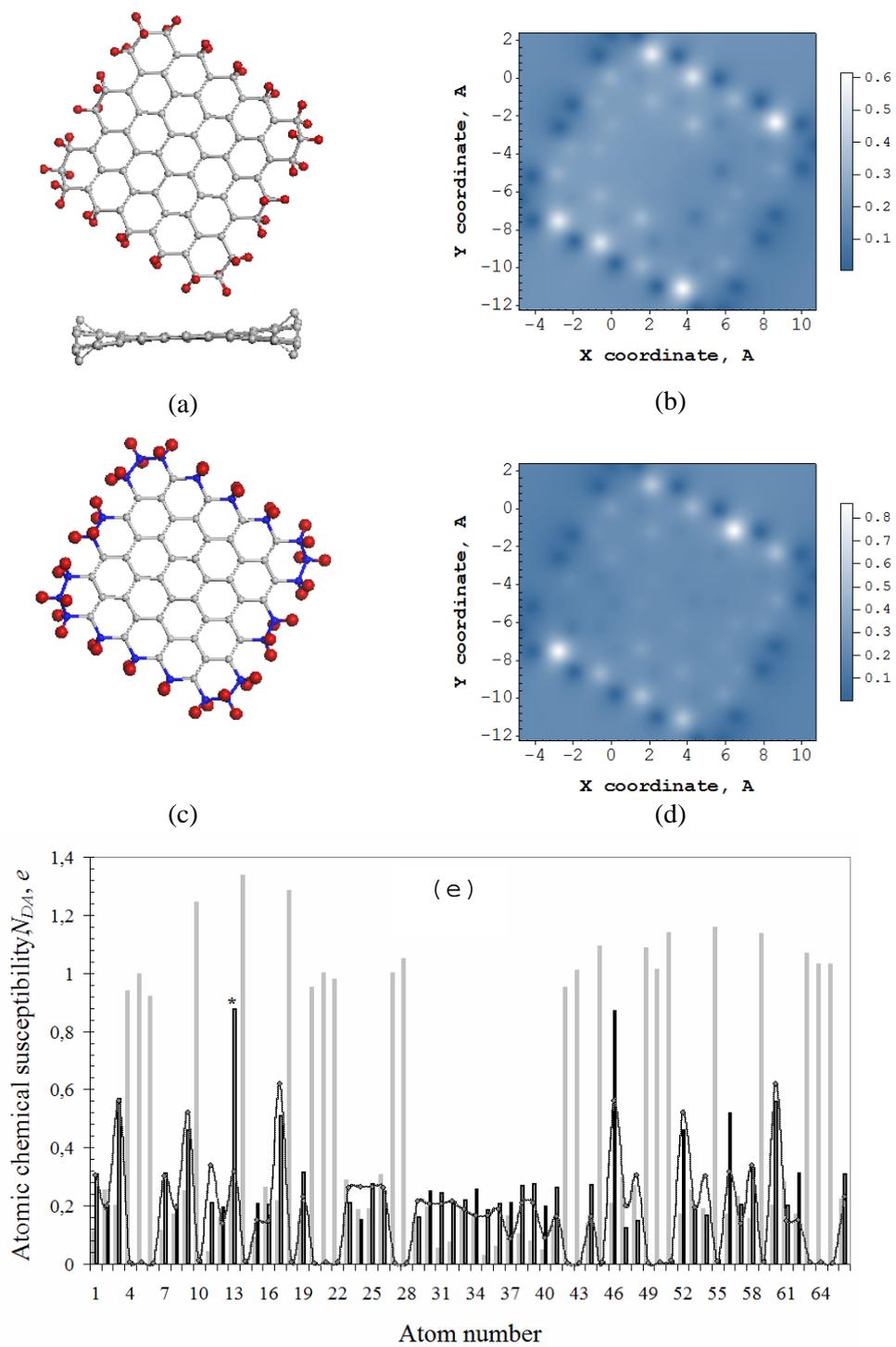

**Figure 2.**

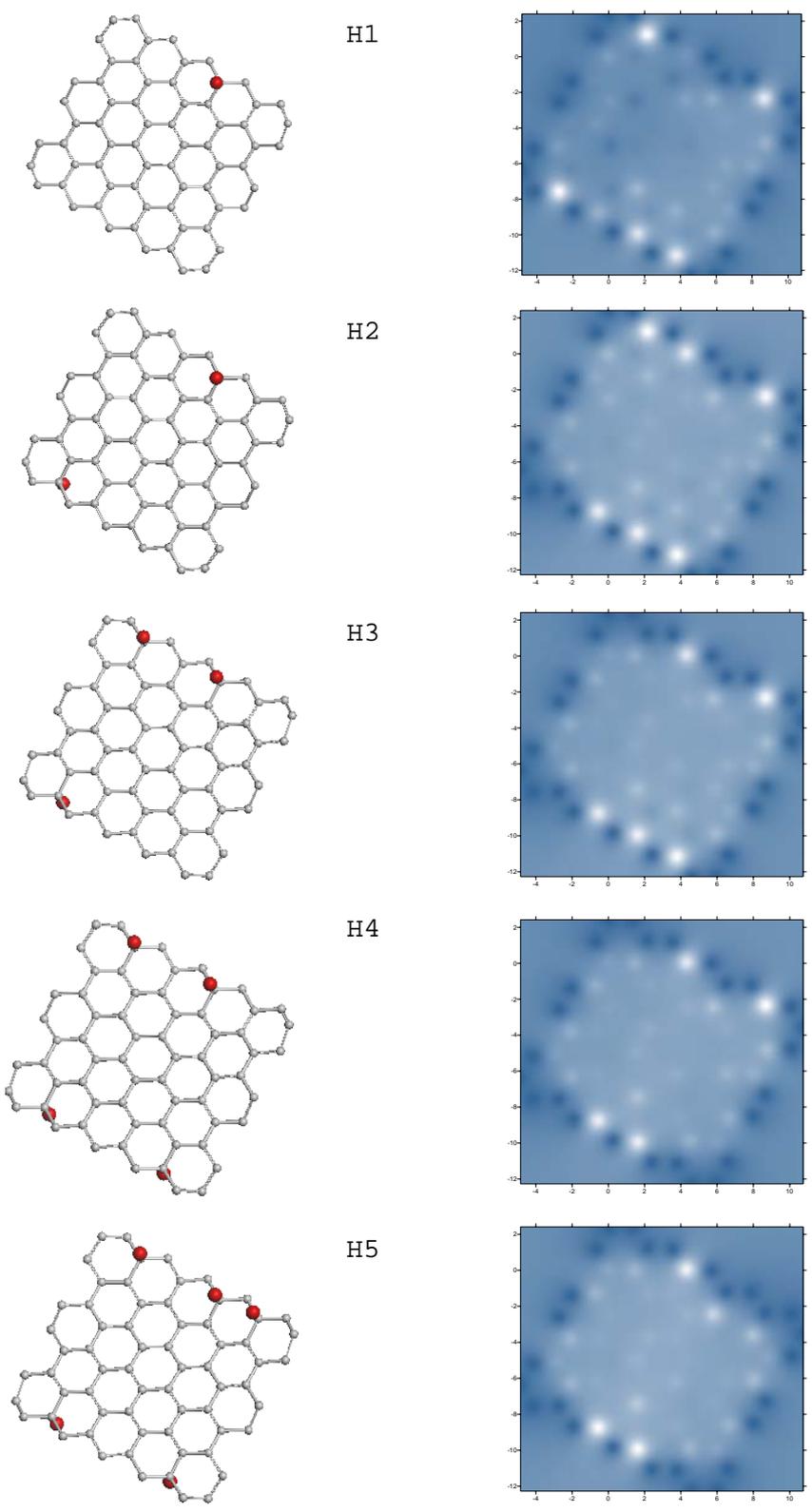

**Figure 3** continued

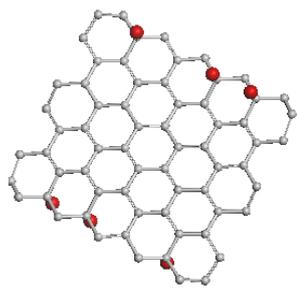 H6 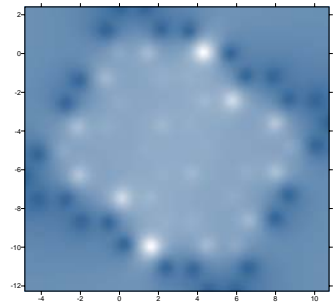

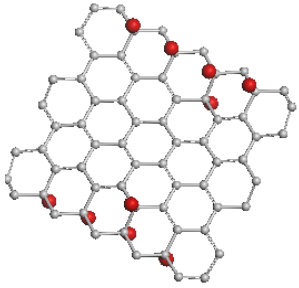 H10 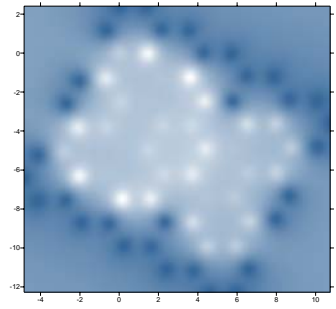

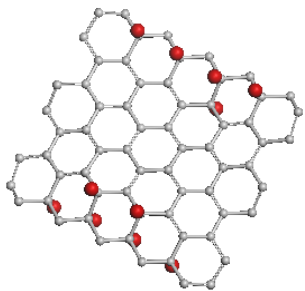 H11 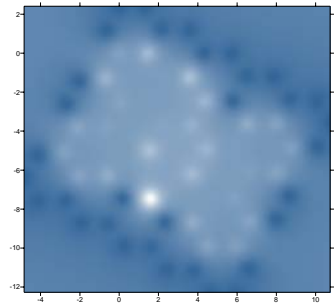

**Figure 3.**

*Top view*    *Side view*

(a) 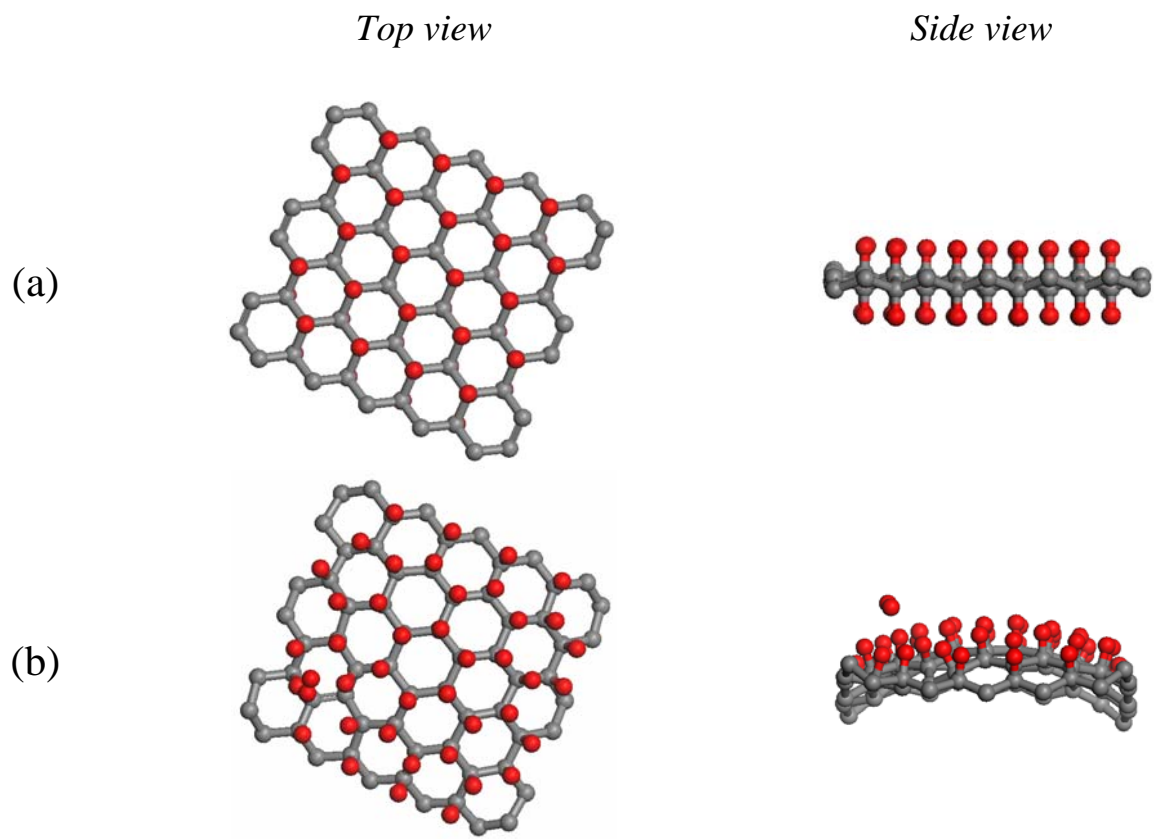

(b)

**Figure 4.**

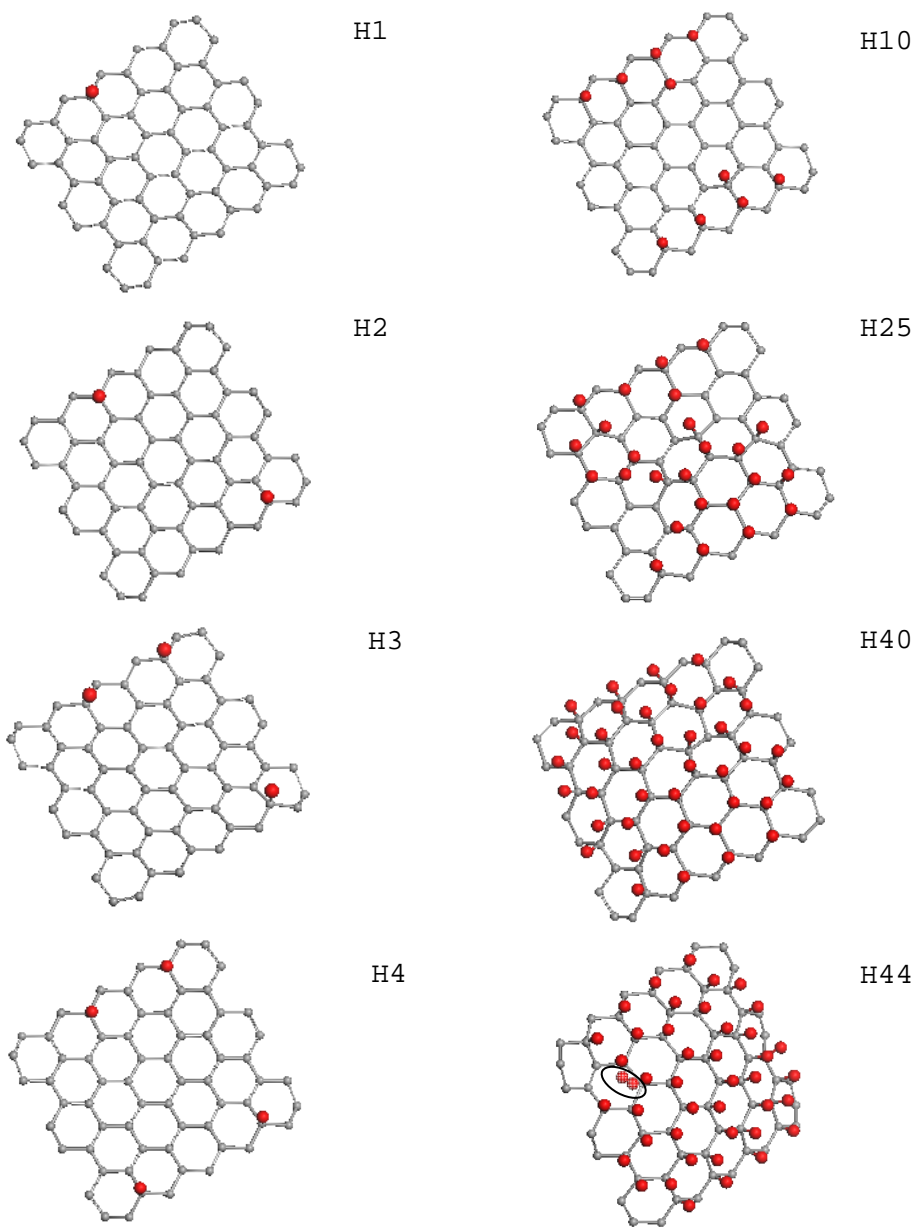

**Figure 5**

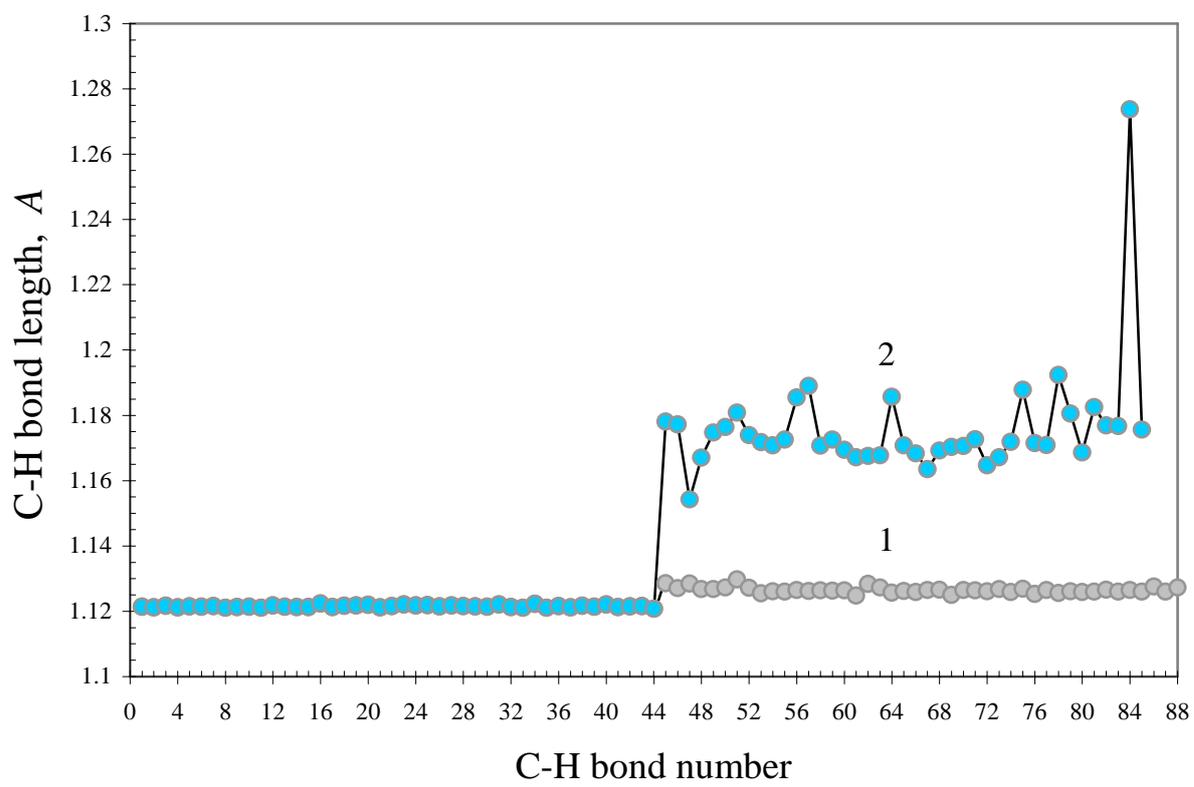

**Figure 6**

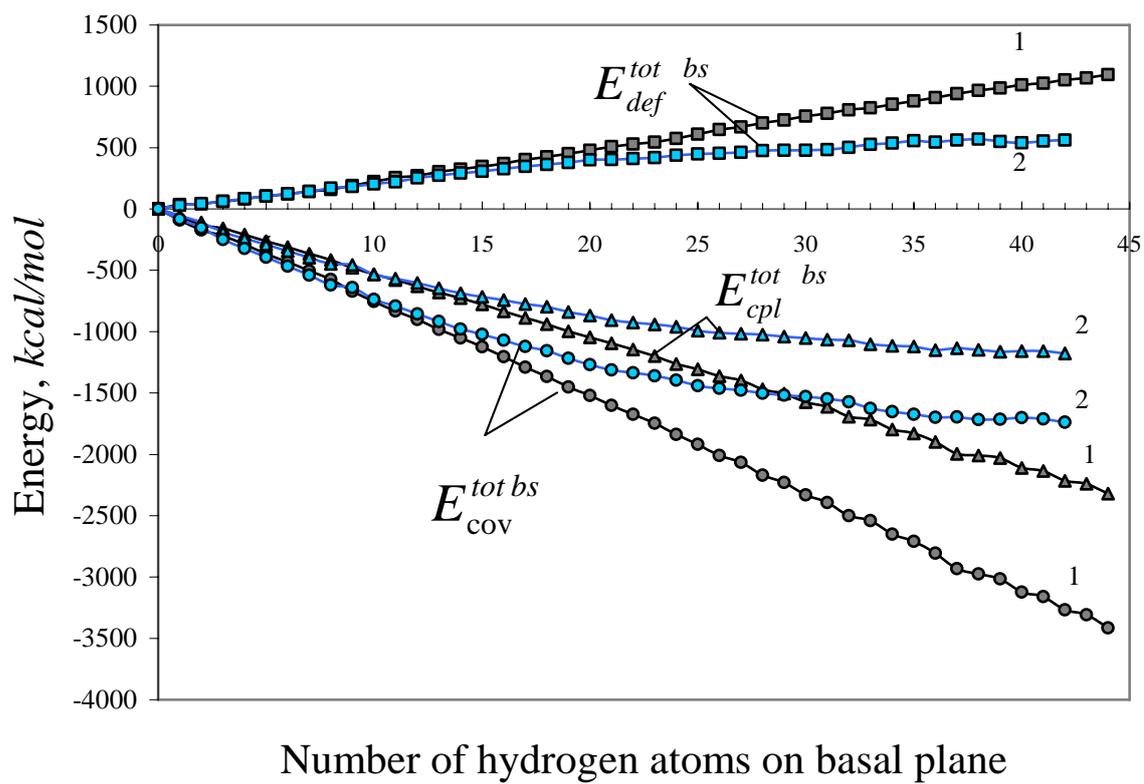

**Figure 7**

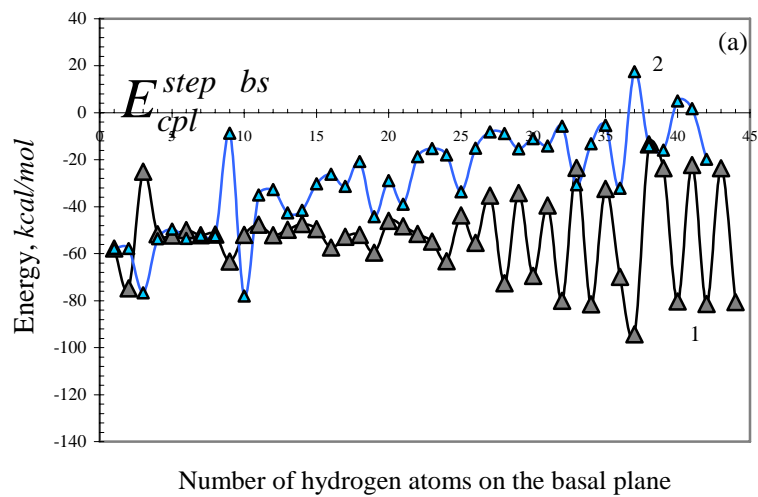

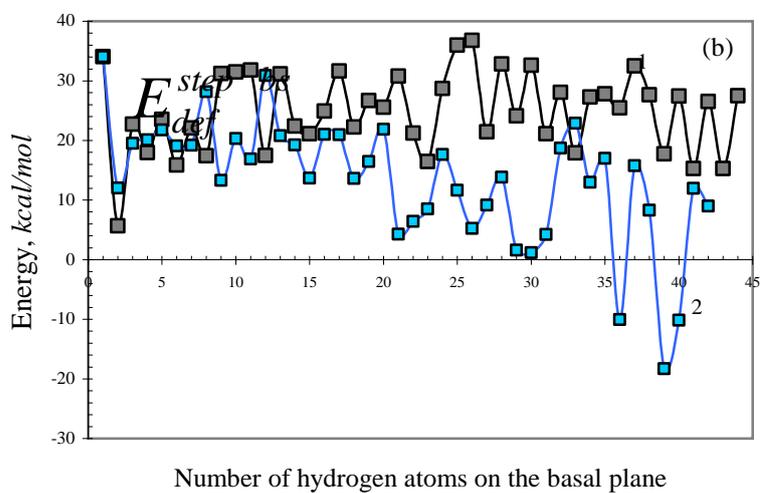

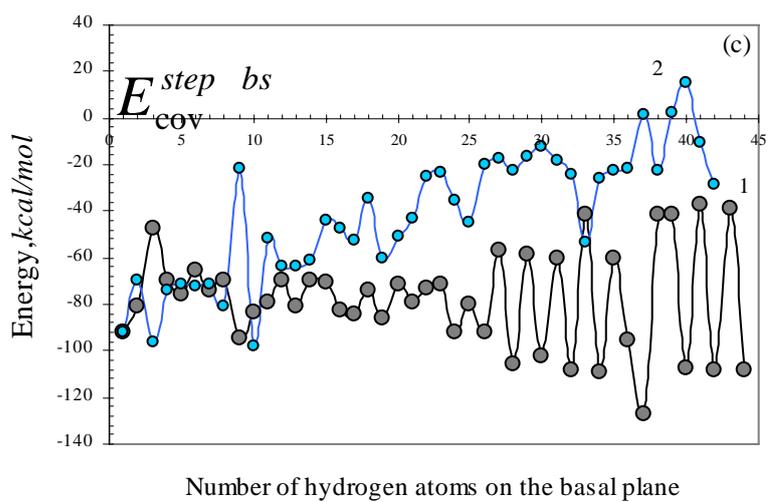

**Figure 8.**

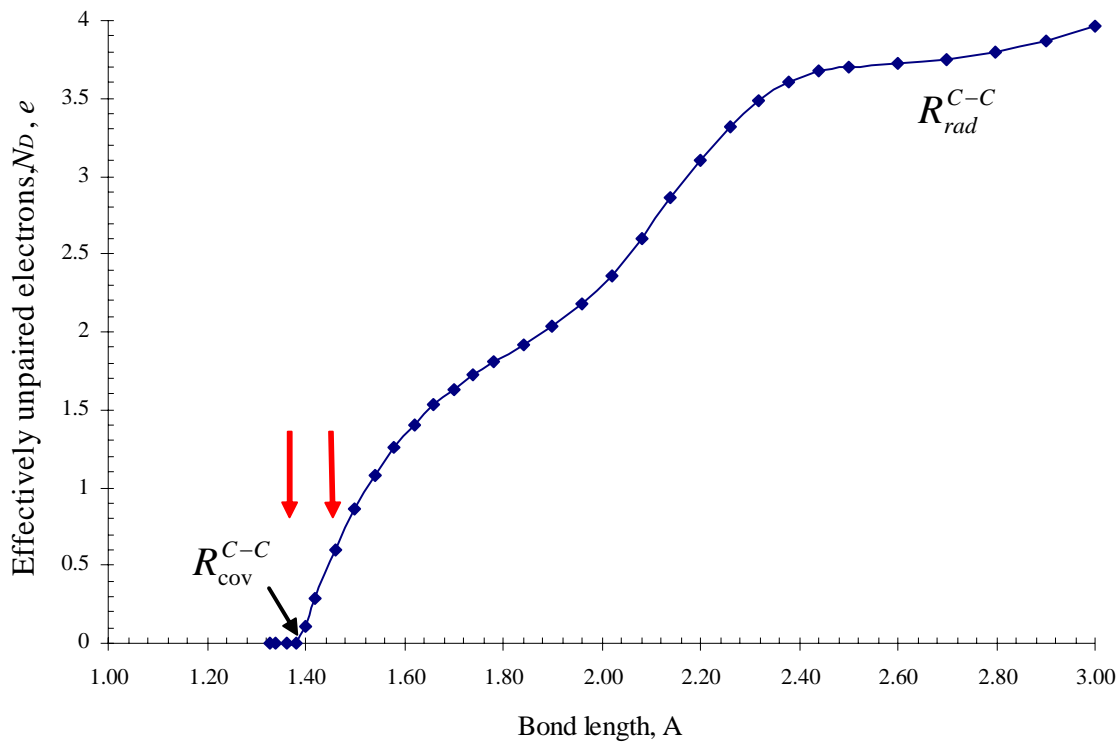

**Figure 9.**

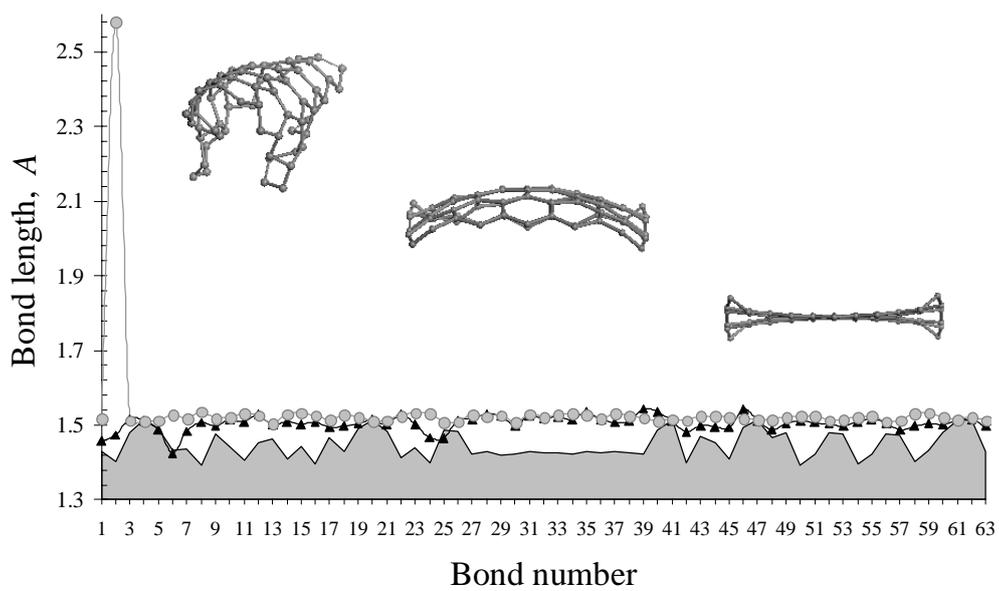

**Figure 10.**

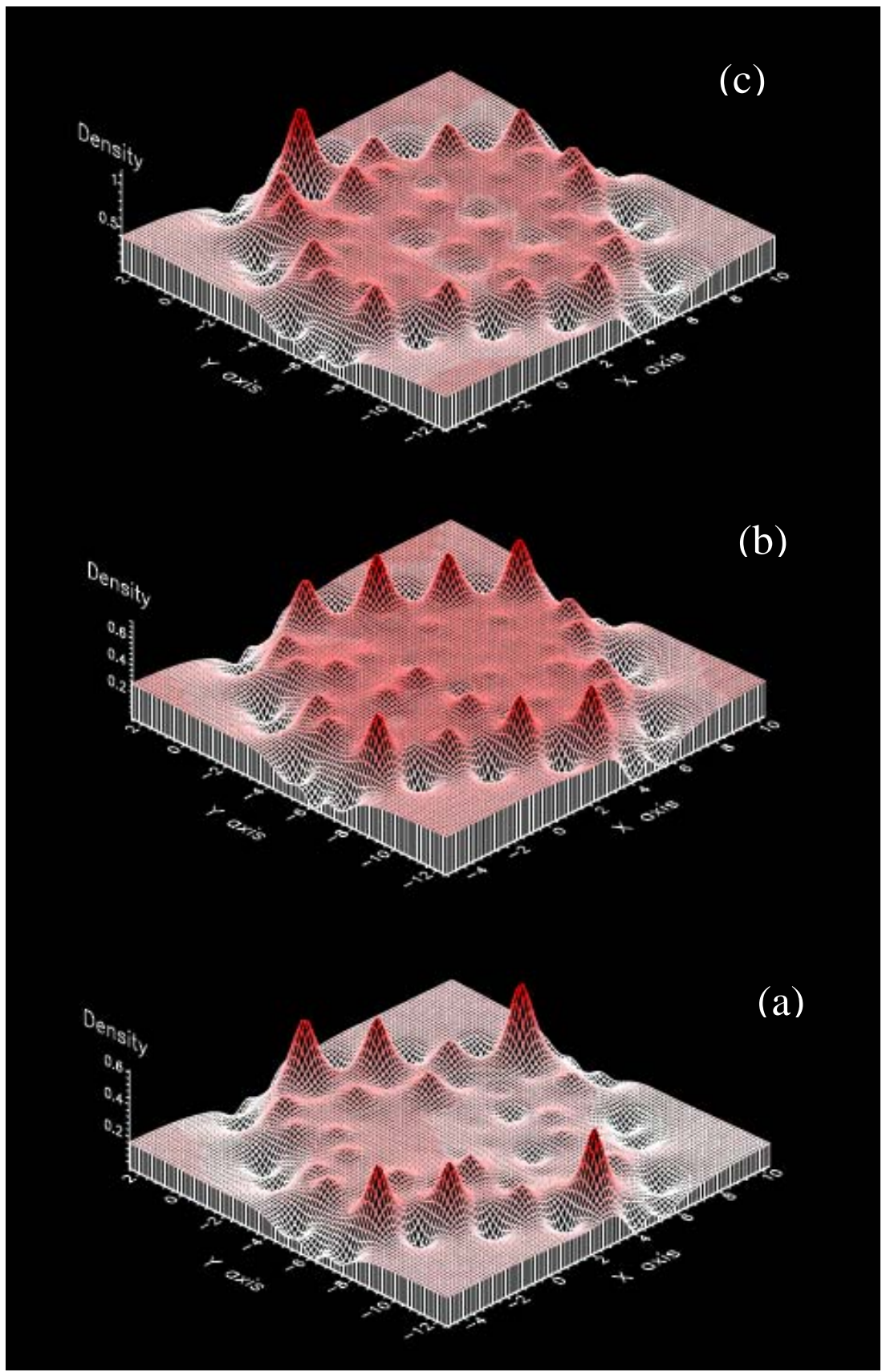

**Figure 11**